\documentclass{INTERSPEECH2023}

\usepackage{multirow}
\usepackage{makecell}
\interspeechcameraready

\title{Progressive Sub-Graph Clustering Algorithm for Semi-Supervised Domain Adaptation Speaker Verification}

\name{Zhuo Li$^{1,2}$, Jingze Lu$^{1,2}$, Zhenduo Zhao$^{1,2}$, Wenchao Wang$^{1}$, Pengyuan Zhang$^{1,2}$}

\address{
  $^1$ Key Laboratory of Speech Acoustics and Content Understanding, \\ Institute of Acoustics, Chinese Academy of Sciences, Beijing, China \\
  $^2$University of Chinese Academy of Sciences, Beijing, China }
\email{li\_zhuo@foxmail.com}

\begin{document}

\maketitle
 
\begin{abstract}
Utilizing the large-scale unlabeled data from the target domain via pseudo-label clustering algorithms is an important approach for addressing domain adaptation problems in speaker verification tasks.
In this paper, we propose a novel progressive subgraph clustering algorithm based on multi-model voting and double-Gaussian based assessment (PGMVG clustering). 
To fully exploit the relationships among utterances and the complementarity among multiple models, our method constructs multiple $k$-nearest neighbors graphs based on diverse models and generates high-confidence edges using a voting mechanism.
Further, to maximize the intra-class diversity, the connected subgraph is utilized to obtain the initial pseudo-labels. Finally, to prevent disastrous clustering results, we adopt an iterative approach that progressively increases $k$ and employs a double-Gaussian based assessment algorithm to decide whether merging sub-classes.

\end{abstract}


\section{Introduction}

Speaker verification (SV) aims to verify the identities of speakers
from samples of their voices, and has been successfully deployed in many
commercial applications. 
State-of-the-art SV systems typically consist of two parts: a frontend for extracting speaker embeddings, and a backend for scoring. For the frontend, deep neural network models such as X-vectors \cite{xvectorformal}, ECAPA \cite{ecapa}, ResNet \cite{ResNet,liming,8759958} are commonly used to extract speaker embeddings. As for the backend, probabilistic linear discrimination analysis (PLDA) \cite{plda00,plda02} and cosine similarity scoring are commonly used for scoring.

In recent years, various large-scale datasets have greatly promote the development of SV systems, such as 
VoxCeleb \cite{vox1,vox2,cslvox}, CTS superset \cite{cts}, CNCeleb \cite{fan2020cn,li2022cn} and so on.
However, the significant domain mismatch between these training data and real-world test data results in a considerable performance degradation of SV systems. In practical applications, a large amount of unlabeled target domain data is easily collected, while manual annotation is labor-intensive and time-consuming.
Therefore, fully utilizing a large amount of  unlabeled target domain data is crucial to boost the performance of SV in practial applications.

To eliminate the domain mismatch, unsupervised domain adaptation methods \cite{wilson2020survey} are proposed in early years.
The commonly used unsupervised adaptation can be mainly 
divided into two categories, one is using adversarial training strategy to learning 
domain-invariant speaker embeddings \cite{wang2018unsupervised,fang2019channel,rohdin2019speaker,xia2019cross,chen2020cross},  the other is  aligning statistic of speaker  embeddings extracted from training and test data \cite{sun2017correlation,lee2019coral+}.
However, these methods only utilize the data distribution of different domains, while other easy-to-access information of target domain is not leveraged.
\cite{chen2021self} introduces a self-supervised learning based domain adaptation method, which exploits the correlation of the target domain data, but the data utilization is still insufficient.

To mitigate the adverse impact of domain mismatch, generating pseudo-labels of unlabeled target data for supervised training has proven to be an effective approach.  Several studies have utilized different clustering algorithms to estimate pseudo-labels, including agglomerative hierarchical clustering (AHC) algorithm \cite{mccree2014unsupervised,thienpondt2020idlab}, autoencoder-based semi-supervised curriculum learning scheme based on Hierarchical Clustering \cite{zheng2019autoencoder}, and k-means algorithm \cite{cai2021iterative,duke,sjtu,slavivcek2021phonexia}. However, AHC is not suitable for large-scale datasets, and the curriculum learning scheme is time-consuming. Although k-means is simple and effective for large-scale datasets, it is sensitive to the number of clusters and not robust to class imbalance. 
Further, many studies have shown that different models can be highly complementary. Traditional clustering algorithms fail to fully exploit this advantage, potentially resulting in suboptimal performance.                                                         
%thereby potentially limiting their performance.

In this study, we propose a novel clustering algorithm, a progressive sub-graph clustering algorithm  based on multi-model voting and double-Gaussian based assessment (PGMVG clustering).
Utterances from the same speaker tend to have smaller distances than those from different speakers in the speaker embedding space. Thus, it can be inferred that an utterance is highly likely to belong to the same speaker as its $k$ nearest neighbors.
Hence, we utilize $k$-nearest neighbors ($k$NN) to construct an Instance Pivot Subgraph (IPS) for each utterance, which the utterance served as a pivot instance and neighboring utterances served as nodes. The $k$-nearest neighbors are connected to the pivot by edges to form the subgraph.

Secondly, numerous studies demonstrate that utilizing results from multiple SV systems can significantly enhance performance. Since SV systems with different structures, initialization or
training schemes tend to extract different speaker features under the constraints of different loss functions, it is crucial to leverage the complementarity between them effectively.
Therefore, in our proposed algorithm, we construct multiple Instance Pivot Subgraphs for all utterances using multiple SV systems. Then, we employ a voting mechanism to select high-confidence edges and construct a multi-model Instance Pivot Subgraph (MIPS). Specifically, we consider the two utterances connected by an edge belong to the same speaker in MIPS.

Subsequently, we aggregate the MIPS of all utterances to obtain the initial speaker relationship graph under the initial $k$ value. Within this graph, all utterances belonging to the same connected sub-graph are assigned with the same pseudo label to preserve the richness of the intra-class distribution to the greatest extent possible. Meanwhile, isolated nodes, which have no edges directly connecting it to any other node, are temporarily removed.
Notably, due to the sensitivity of the connected sub-graph search method to false positive edges, setting an optimal $k$ value directly is non-trivial. Therefore, we address this issue by progressively increasing $k$.

As $k$ progressively increases, new edges and nodes are added in speaker relationship graph,
potentially resulting in the merging of current pseudo labels. 
While this process may be accompanied by adding new false positive edges.
We construct trials by grouping some utterances and score these trials, 
the scores follow a double Gaussian distribution if these utterances come from different speakers, 
with one Gaussian representing the distribution of positive trials 
scores and the other representing the distribution of negative trials scores.
Based on this finding, we introduce the double-Gaussian based assessment, which utilizes the double Gaussian distribution to fit the scores of trials constructed by utterances to assess whether utterances belong to the same speaker, and thus decide whether to merge pseudo labels and retain added edges \& nodes.
Experimental results demonstrate the superiority of our method over others.

%Based on this finding, we introduce the double-Gaussian based assessment, which utilizes the double Gaussian distribution to fit the scores of trials constructed by utterances to assess whether utterances belong to the same speaker.
%Experimental results demonstrate the superiority of our method over others.

\section{Method}

\subsection{Algorithm Objective}
The labeled source domain data and 
unlabeled target domain data are denoted as $S$ and $T$, 
respectively.
Suppose that there are $N$ speaker feature extractors $E_n$, where $n \in \{1, 2, ..., N\}$, 
trained with supervised paradigm using labeled source domain data $S$.
Then, speaker embeddings $x_{m,n}$ of utterances $t_m,\ m \in \{1, 2, ..., M\}$, from unlabeled taget domain are extracted by $E_n$.
%All utterances $t_m$, $i \in \{1, 2, ..., M\}$, from the unlabeled target 
%domain $T$ are processed by $E_n$ to extract speaker embeddings $x_{m,n}$.
The goal of the algorithm is to assign a pseudo label $y_{m}$
to each $t_m$ using $x_{m,n}$, thereby re-training the speaker feature extractors using the unlabeled utterances from the target domain.

\begin{figure}[!th]
%\vspace{-2mm}
\centering
\includegraphics[width=1.0\linewidth]{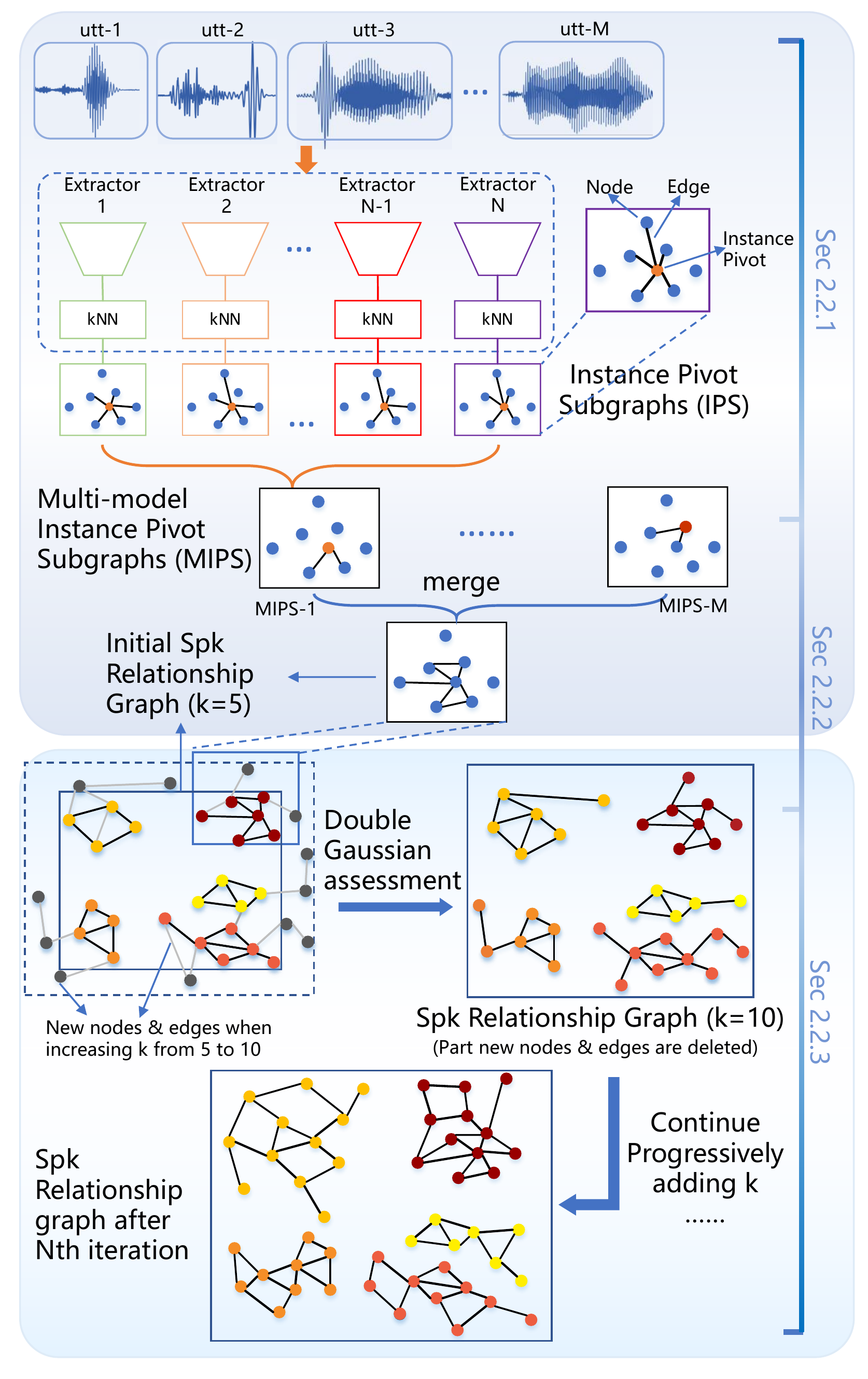}
\vspace{-4mm}
\caption{Pipeline of the progressive sub-graph clustering algorithm based
on multi-model voting and double-Gaussian based assessment.}
\label{fig}
\vspace{-4mm}
\end{figure}

\subsection{Algorithm Pipeline}

\subsubsection{Multi-model Instance Pivot Subgraphs Construction}

For all utterances in the target domain $t_m$, speaker embeddings are first extracted using multiple speaker feature extractor, denoted as $x_{m,n}$. 
Then, IPS$_{m,n,k}$ is constructed for each utterance $t_m$ by using $x_{m,n}$ and searching $k$-most similar utterances to $x_{m,n}$ using $k$NN algorithm.
As shown in Fig.~\ref{fig}, multiple IPS$_{m,n,k}$ are constructed based on different extractors $E_n$ for each utterance $t_m$.
Next, we use a multi-model voting strategy 
to combine IPS$_{m,*,k}$, 
and obtain multi-model IPS, denoted as MIPS$_{m,k}$.
For simplicity, we only remain edges that exist in all IPS$_{m,*,k}$ in MIPS$_{m,k}$.
Fig.~\ref{fig} shows the MIPS$_{m,k}$ based on the provided IPS$_{m,n,k}$ when $k$ is set as 5.

\subsubsection{Initial Pseudo-labeling}

After obtaining MIPS$_{m,k}$ for all utterances, we construct the initial speaker relationship graph by aggregating them, as depicted in Figure~\ref{fig}. The aggregation principle states that if two utterances (node) are connected by an edge in any IPS, they are connected in the speaker relationship graph. We then obtain initial pseudo-labels by identifying connected sub-graphs, and assigning the same pseudo-label $y_{m}$ to all utterances in a sub-graph. Sub-graphs with less than ten utterances are excluded from pseudo-label assignment. In addition, isolated utterances, which have no edges connecting to other utterances, and utterances in sub-graphs that contains less than ten utterances are removed from the current speaker relationship graph.

\subsubsection{Progressive Pseudo-Labeling Iteration}

After obtaining the initial pseudo-labels under the 
initial k-value, we progressively increase $k$ 
to add new edges and nodes to the current graph.
New edges can be classified into three types depending on whether the connected utterances are already existing in the current speaker relationship graph.

For convenience, we adopt the notation $X_{gin}$ to denote the set of utterances already existing in the current speaker relationship graph and $X_{gout}$ to represent the set of utterances not yet included in the current graph.
Suppose that $t_a$ is one node connected by the new edge, 
and $t_b$ is the other one.

The following are descriptions of the three cases: (1) $t_a \in X_{gin}$ and $t_b \in X_{gin}$;
(2) $t_a \in X_{gout}$ and $t_b \in X_{gout}$;
(3) $t_a \in X_{gin}$ and $t_b \in X_{gout}$.
In the proposed algorithm, principles are designed 
to determine whether keep the new edges or nodes in these 
three cases, sequentially.
For CASE (1), if $t_a$ and $t_b$ are originally given the same pseudo-label, keep this edge.
If they have different pseudo-labels, 
a double-Gaussian based Assessment is proposed to decide 
whether merging the two sub-classes or not.
The double-Gaussian based Assessment is
is described in detail in the subsection \ref{sd}.
For CASE (2), 
directly search the connected sub-graph 
for the involved nodes, without detecting edges.
For CASE (3), 
when the $t_b \in X_{gout}$ connects to multiple sub-classes, 
we introduce double-Gaussian based Assessment to decide whether 
the sub-classes need to be merged. If not, delete this node.

The above algorithms is processed iteratively, as the k-value increases.
The conditions for stopping the algorithm is two-fold.
One is that, the number of newly added nodes 
is less than the an initial value setting, usually 1\% of all nodes.
The other is, the number of sub-classes tends 
to be stable.
The appearance of these two phenomena 
indicates that most of the utterances have already 
been assigned with an appropriate pseudo-label.

\subsubsection{Double-Gaussian based Assessment}% for Sub-Class Merging}
\label{sd}
In this subsection we introduce the double-Gaussian based assessment algorithm to decide whether merging sub-classes.

\begin{figure}[!hb]
\vspace{-2mm}
\centering
\includegraphics[width=1.0\linewidth]{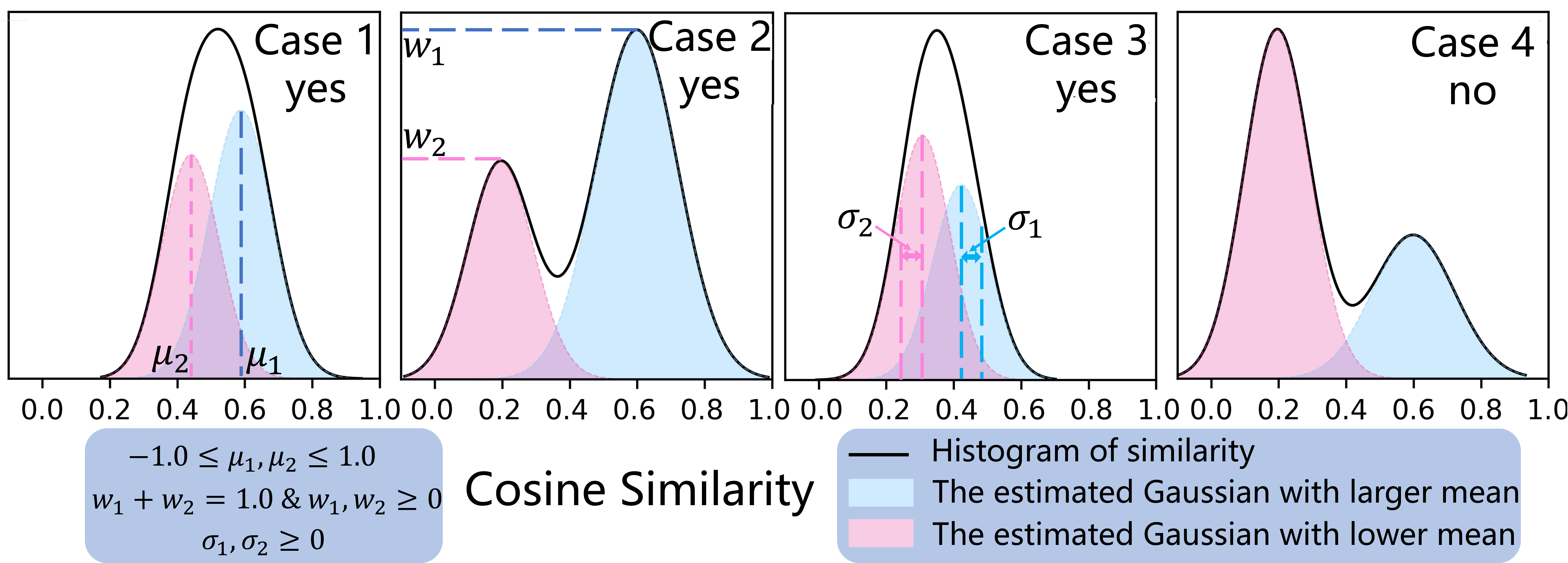}
\vspace{-4mm}
\caption{Double-Gaussian based Assessment}
\label{fig2}
\vspace{-2mm}
\end{figure}
\vspace{-1mm}

For sub-classes that need to determine whether to merge or not,
we calculate the similarity between all $x_{m,n}$ from
these sub-classes for each extractor.
If these sub-classes come from different speaker, 
scores follow a double Gaussian distribution, approximately. Otherwise, scores follow a single Gaussian distribution.
Therefore, we fit the distribution using a double Gaussian model,
and obtain parameters \{$\mu_1, \mu_2, \sigma_1, \sigma_2, w_1, w_2$.\}
For convenience, suppose that $\mu_1 > \mu_2$.
All possible cases could be divided into four categories.
The principles of merging based on double-Gaussian based assessment
in these cases could be concluded as follows, as depicted in Figure~\ref{fig2}:

\begin{enumerate}
  \item If $\mu_2 > th_{high}$, $th_{high}$ 
  is a initial setted threshold, 
  then merge these sub-classes, as shown in Fig.~\ref{fig2} (CASE 1).
  \item When CASE 1 is not satisfied, if $w_1 > 0.5$, 
  then merge these sub-classes,
  as shown in Fig.~\ref{fig2} (CASE 2).
  \item When CASE 1 and 2 are not satisfied, 
  if $\mu_1-\sigma_1<(\mu_2+\sigma_2)+\epsilon$, where $\epsilon$ is a very small value, it indicates a significant overlap between the two Gaussian, then merge these sub-classes, as shown in Fig.~\ref{fig2} (CASE 3).
  \item In other cases, do not merge, 
  as shown in Fig.~\ref{fig2} (CASE 4).
\end{enumerate}

These principles is primarily based on 
whether the number of noisy nodes in the merged sub-classes is 
within an acceptable range.
For CASE 1, when $\mu_2$ is high enough, 
speaker embeddings in these sub-classes 
are extremely similar.
These sub-classes could basically be determined to 
be the same.
For CASE 2, the reason why the merging is available
is that, a moderate amount of noisy samples 
is tolerable.
When $w_1 > 0.5$, after merging, 
a larger number of nodes in the dominant class could be guaranteed.
For case 3, the two Gaussians 
can be basically considered 
as the same Gaussian.
The reason for this situation we attribute 
to inadequate fitting of base models.
We set a constrain $\mu_1>th_{low}$, where $th_{low}$ 
is a lower threshold, to 
prevent too many sub-classes being merged together.
For other conditions, it is highly likely that these sub-classes belong to different speakers.

After assessing these sub-classes with N speaker extractors $E_*$, we obtain N decisions. Then, we combined them to get the final decision by using a voting mechanism, following the principle of majority rule.

\section{Experimental Setup}

\subsection{Dataset and Data Usage}
\noindent \textbf{Training data:}
Similar to the VoxSRC22 competition, we utilize the following datasets as the training data. 

\noindent(1) The VoxCeleb2\cite{vox2,cslvox} dev set with speaker labels is used as the labeled data from the source domain. 

\noindent(2) A subset of the CNCeleb2\cite{li2022cn} dev set without speaker labels is used as the unlabeled data from the target domain, utterances from the “singing”, “play”, “movie”, “advertisement”, and “drama” genres are removed from the CNCeleb2 dev set. 

\noindent (3) A small subset of CNCeleb with speaker labels is used as the labeled data form the target domain. 

\noindent \textbf{Testing data:} We use the official validation and test trial list of the VoxSRC22 Track 3 to evaluate our system, denote it as VoxCN-dev and VoxCN-eval, respectively. 

\noindent \textbf{Data cleansing:}
Because of the lack of filtering when constructing the CN-Celeb2\cite{fan2020cn,li2022cn}, it contains much noisy audio. One of the most intuitive manifestations is that there is much-repeated audio in CN-Celeb2, and some are given different labels.
We directly used $md5sum$ to de-duplicate the speech and the number of audio decreased from 455,946 to 409,628.

\subsection{Model Training and Setting}

\noindent \textbf{Data processing:}
To augment data, we first use the SoX speed function with speeds 0.9 and 1.1 to generate extra twice speakers \cite{speed}. Then, we use MUSAN \cite{musan} and RIRs noises \cite{rirs} to perform online data augmentation. Similar to \cite{zhao2022multi}, a chain augment pipeline is used to generate samples: (1) MUSAN noise with probability 0.2, (2) MUSAN music with probability 0.2, (3) MUSAN speech with probability 0.2, (4) RIRs noises with probability 0.6. 64/80-dimensional Fbank is used as input features without VAD.

\noindent \textbf{Training protocol for base model:}
We selected the following five models as base models:
(1) \emph{ResNet34} with 32 channel and SE module; (2) \emph{CoT-Net}\cite{cot} with 32 channel; (3)\emph{Conformer-MFA} with 256 hidden dim \cite{zhang2022mfa}; (4)\emph{ECAPA-TDNN} with 1024 channel; (5) \emph{SE-ResNet101} with 32 channel; (5) \emph{ResNet101}\cite{cot} with 32 channel and SE module;
Training settings for these base models are shown in Table~\ref{t-base}. Model (1-2) are trained with SGD optimizer and no augmentation is used. The learning rate is set to 0.1, 0.01, 0.001 and is switched when the training loss plateaus. Model (3-5) are trained using data augmentation and a two-stage protocol, cycle learning rate scheduler is adopted in the first stage, details are shown in the latter. 
Models (3-5) are trained using data augmentation and a two-stage protocol that utilizes a cycle learning rate scheduler in the first stage. The details of the protocol are shown in the subsequent paragraph. Other setting are shown in Table\ref{t-base}. These five models are all trained with circle loss.

%============== adding

\noindent \textbf{Training protocol for final model:}
After obtaining pesudo labels of unlabeled data, we train final models with a two-stage protocol. In the first stage, the SGD optimizer with a momentum of 0.9 is used and weight decay of 2e-4 is used. Mini-batch size is 1024 and the segment duration is 2s. ReduceLROnPlateau scheduler is adopted, where the initial learning rate is 0.1, the minimum learning rate is 1e-6, and the decay factor is 0.1 or 0.5. 
Circle loss with Subcenter and Intertopk is adopted, with hyperparameters set as follows: s=60, m=0.35 for Circle loss, k=3 for Subcenter, and k=5, m=0.1 for Intertopk. In addition, we explore two training strategies in this stage, one is train the model form scratch, and the other is to utilize models trained with Vox2 as the pre-trained model. For the latter, we freeze the extractor and train the classification layer before start training the whole model. Experiments show that performance of these two methods are similar.

In the second stage, we finetune the model on the CNCeleb dataset using the Adam optimizer with weight decay of 4e-4, while expanding the segment duration to 6s and removing the Intertopk. Specially, to prevent overfitting, we preserve the VoxCeleb weights of the classification layer.

\noindent \textbf{Back-end:}
Cosine similarity is used for matrix scoring\cite{cslvox}, AS-norm and QMFs are used to calibrate the scores.

\subsection{Implementation Details of PGMVG Clustering}
Firstly, we align the centers of different domains based on base models 
to eliminate the domain mismatch, denote it as statistic adaptation.
\footnote{PLDA adaptation methods achieves better performance than statistic adaptation, scores obtained from PLDA adaptation methods exhibit a large range of variability and significant differences across different models, making it challenging to determine a stable empirical value.In addition, PLDA is unable to perform fast matrix operations, whereas cosine similarity can.Therefore, we do not use the PLDA back-end.}
The reason why we adopt it is two-fold. One is that statistic adaptation
is a commonly used method to alleviate domain adaptation problems in SV field, which could 
be considered as a baseline. The other is that using statistic adaptation is able to enhance the effectiveness of clustering. 

Secondly, before constructing the IPS for each speech segment, we remove some speech segments by comparing their cosine similarity with the 500th most similar speech segment. If the similarity score is greater than 0.8, we remove the speech segment. The removed speech segments can be roughly classified into two categories. 
The first category includes speech segments containing music or singing, resulting in an overall higher similarity score with other speech segments and disastrous clustering results. The second category includes speech segments belonging to classes with a good fit or having a large number of intra-class speech segments, and removing them does not significantly affect the performance.The final list of speech segments to be removed was obtained by taking the union of the speech segment lists generated by multiple models.

To obtain high confidence edges by using voting strategy, it is beneficial to select models
with as much variance as possible, from the structure to the training protocol, as shown in Table~\ref{t-base}.
Finally, the initial $k$ is set to 5 and $k$ is added by 5. $th_{high}$ and $th_{low}$ are set to 0.4 and 0.2 for double-Gaussian based assessment, respectively. After clustering algorithm, we obtain 1711 speakers and 348861 speech segments.

\section{Result and Analysis}

We firstly analyze the impact of domain adaptation methods. 
Results before/after adaptation are shown in Table~\ref{t-base}. There are two things worth noting. Firstly, the larger the model, the better its generalization performance, and ResNet101 achieved the optimal performance. Secondly, using statistical adaptation leads to a stable performance improvement of around 15\%.
\begin{table}[htbp]
  \centering
  \caption{Results of base models before/after adaptation.}
\tabcolsep0.068in
    \begin{tabular}{cccc}
    \toprule
    \multirow{2}[2]{*}{System} & \multirow{2}[2]{*}{Training set} & \multicolumn{2}{c}{VoxCN-dev-EER} \\
           &       & before   & after \\
    \midrule
    ResNet34 &  Vox2-clean-Fb64-SGD & 16.86 & 14.29 \\
    CoT-Net &  Vox2-clean-Fb64-SGD & 16.65 &  14.55\\
    \midrule
    Conformer & Vox2-aug-Fb80-Adam & 16.95 & 14.14 \\
    ECAPA-large & Vox2-aug-Fb80-Adam & 18.02 & 14.62 \\
    ResNet101 & Vox2-aug-Fb80-Adam & 14.06 & 11.90 \\
    \bottomrule
    \end{tabular}%
  \label{t-base}%
\end{table}%

Subsequently, we train three models, ResNet34, Res2Net50 and PCF-ECAPA\cite{zhao2023pcf}, using both the unlabeled target domain data that has been assigned pseudo-labels through our proposed method and all the labeled data. Results are shown in Table~\ref{t2}.
The performance of Res2Net and ResNet is significantly better than that of PCF-ECAPA, which we conjecture is due to the stronger modeling ability of the former two, making them less susceptible to the the erroneous labels in the pseudo-labels. Of course, they all achieve better performance than domain adaptation methods, as they leverage unlabeled data more effectively.

\begin{table}[htbp]
  \centering
  \caption{EER of various systems on the VoxCN-dev.}
\tabcolsep0.068in
    \begin{tabular}{ccccccc}
    \toprule
   \makecell[c]{Clustering \\ method}       &   No.   & System &       & \makecell[c]{$without$ \\calib} &       & \makecell[c]{$with$ \\ calib} \\
\cmidrule{1-3}\cmidrule{5-5}\cmidrule{7-7}    \multirow{3}[2]{*}{\makecell[c]{PGMVG}} & S1    & ResNet34 &       & 8.61  &       & 7.66 \\
          & S2    & Res2Net50 &       & 8.45  &       & 8.20 \\
          & S3    & PCF-ECAPA &       & 9.57  &       & 8.81 \\
\cmidrule{1-3}\cmidrule{5-5}\cmidrule{7-7}    \multirow{2}[2]{*}{\makecell[c]{k-means}} &       & ResNet34\cite{duke} &       & 9.73     &       & 8.45 \\
          &       & ResNet34\cite{sjtu} &       & -- &       & 8.82 \\
    \bottomrule
    \end{tabular}%
  \label{t2}%
\end{table}%

Furthermore, compared to using k-means for pseudo-label generation, our algorithm achieves better results, as it maximally preserved the intra-class diversity. When the number of speakers in the training set is similar, increasing the intra-class diversity leads to better performance gains.
Finally, we present the results of our fusion system on the VoxCN-dev \& eval sets, and compare them with other state-of-the-art (SOTA) results in Table~\ref{t3}, which further validates the effectiveness of our method.

\begin{table}[htbp]
  \centering
  \caption{Results of fusion systems on the VoxCN-dev \& eval.}
    \begin{tabular}{cccccc}
    \toprule
    \multirow{2}[4]{*}{System} & \multicolumn{2}{c}{VoxCN-dev} &       & \multicolumn{2}{c}{VoxCN-eval} \\
\cmidrule{2-3}\cmidrule{5-6}          & EER   & minDCF &       & EER   & minDCF \\
    \midrule
    S1+S2+S3 (ours) & 6.77  &  0.298  &       & 7.03  & 0.388 \\
    Fusion \cite{duke} & 7.00     & 0.326 &       & 7.15 & 0.389 \\
    Fusion \cite{sjtu} & 7.13 & 0.329 &       & 8.08 & 0.437 \\
    \bottomrule
    \end{tabular}%
  \label{t3}%
\end{table}%

\section{Conclusions}
In this paper, we propose a novel clustering algorithm, named PGMVG clustering, to fully leverage the large amount of unlabeled data in the target domain for domain adaptation. Our algorithm introduces a voting mechanism to effectively utilize information from multiple models, and utilizes connected subgraphs to enhance intra-cluster diversity as much as possible. Furthermore, a double-Gaussian based assessment algorithm and a progressive strategy are introduced to prevent disastrous results. Experimental results demonstrate the effectiveness of our method.
Additionally, our algorithm requires further exploration and optimization in several aspects. First, using multiple empirical parameters may pose challenges for novice users and can benefit from the addition of score calibration. Second, label correction is crucial for the algorithm and will be explored in future work. Finally, an important research direction is to optimize the algorithm for speaker diarization tasks.

\bibliographystyle{IEEEtran}
\bibliography{mybib}

\end{document}